\documentclass[aps,prl,twocolumn,showpacs,floatfix]{revtex4}
\usepackage{graphicx}
\usepackage{bm,amsmath,amssymb}
\newcommand{\bos}{\boldsymbol}

\begin{document}
\title{Block copolymer knots }
\author{Franco Ferrari$^1$}
\email{franco@feynman.fiz.univ.szczecin.pl}
\affiliation{$^1$CASA* and Institute of Physics, University of Szczecin,
  Szczecin, Poland} 
\date{\today}

\begin{abstract}
  An extensive study of single block copolymer knots   containing two kinds of
monomers $A$ and $B$ is presented. The knots are in a solution and
their monomers are subjected to short-range interactions that can be
attractive or repulsive. 
In view of possible  applications 
 in medicine and the construction of
intelligent materials, it is shown that several features
of copolymer knots
can be tuned by changing the monomer configuration.
A very fast and abrupt swelling with increasing 
temperature is
obtained in certain multiblock copolymers, while the size and the swelling
behavior at high temperatures may be controlled in diblock
copolymers. Interesting new  effects  appear in the thermal diagrams of
copolymer knots when their length is increased.

\end{abstract}
\maketitle

Polymer knots are abundant in nature and in artificial polymer
materials \cite{sauvage,arsuaga,ramirez,amabilino,tezuka}. They can be created
in the laboratory \cite{amabilino,tezuka} and have attracted a 
considerable attention both from experimentalists and theoreticians of
several different disciplines. When direct measurements are too difficult
to be performed, numerical simulations are used to provide reliable
predictions on the properties of these knots.
So far, an important aspect of polymer knots has not been studied or
studied marginally~\footnote{In Ref.~\cite{wang} a particular kind of
  multiblock copolymers with a fixed length of forty segments has been
  investigated, which  does not sustain the remarkable properties
  found in the   longer copolymers studied here.},
namely that of block copolymer knots.
This is the goal of the present
work. The copolymer 
knots are defined here on a simple cubic lattice and
contain only two kind of monomers, $A$ and $B$.
Monomers of the same kind repel themselves, while the interactions
between $A$ and $B$ are attractive.
 Two classes of copolymer knots will be studied, namely 
the $AB-$diblock 
copolymers and the multiblock copolymers.
These classes  will be called $D(N_A,N_B)$ and $M(s_A,s_B)$
respectively. 
Here $N_A$ and $N_B$ denote  the number of $A$ 
and $B$ monomers forming the two blocks of a $AB-$diblock 
copolymer. Of course $N_A+N_B=N$, $N$ being the total monomer
number. In $M(s_A,s_B)$ copolymers,
a block consisting of
$s_A$ 
monomers of type $A$  followed by $s_B$ monomers of type $B$, is
repeated $n-$times. For 
example, $M(2,2)$ corresponds to 
$(A_2B_2)_n$ multiblock copolymers.
Furthermore, 
homopolymer knots with $N$
monomers subjected to short attractive and
repulsive interactions will be denoted with 
the symbols  $H(N,-)$ and $H(N,+)$ respectively.

The rationale for investigating such kind
of copolymers is to obtain polymer knots with
different behaviors by changing the monomer distribution.
Particularly suitable to this purpose is the  class of $D(N_A,N_B)$
copolymers. In this class the conformations 
in which  the monomers $A$ and $B$ are close
to each others are energetically favored because they have a lower energy
due to the attractive
interactions between the two different types of monomers.
On the other side, the $A$ and $B$ monomers are concentrated in
two distinct parts of the knot.
As a consequence, for entropic reasons these low energy conformations
are far less 
probable  than those in which the $A$ and $B$ monomers are mostly 
in contact only with themselves, especially when one kind
of monomers is much more abundant than the other. The upshot is that
it is possible to  change the properties of 
$D(N_A,N_B)$ copolymer knots
by
varying the $A/B$ monomer ratio $\eta=\frac{N_A}{N_B}$.
In particular, in view of possible medical applications or of intelligent
polymer materials containing knots, we show that the size of the knot
can be tuned to a high extent. More precisely, it is possible to create
polymers whose size 
is approximately stable over a given interval of temperatures, but changes
after some threshold is reached. While homopolymer knots subjected to
short-range interactions are essentially two-regime polymers, undergoing a slow
transition 
from the compact to the swollen phase, at high values of $\eta$ the
$D(N_A,N_B)$ copolymer knots  
admit three regimes.  The compact regime appears at low
temperatures and is characterized by compact conformations.
The attractive interactions are dominant. At intermediate temperatures,
the repulsive interactions prevail and the copolymer is reaching its
maximal extension, so that this regime can be called ultra swollen. At
very high temperatures the interactions cease to be relevant and 
an entropy dominated regime begins, in which the knot is swollen.
Multiblock copolymers in the class $M(s_A,s_B)$ have remarkable
properties too.  
For example,  they exhibit a
transition from the compact to the swollen phase which is abrupt
 in comparison to that of homopolymers and diblock copolymers.
Finally, the phase diagram of
copolymer knots of type $D(N_A,N_B)$ becomes more comples with
increasing values of $N$ as new peaks appear in the heat capacity. 

The used methodology will be now briefly explained. The monomers are located on
the lattice sites in such a way that two consecutive monomers are
linked together by one lattice bond. The bond length is
one, so that the length of the knot is equal to $N$.
The energy  of a given knot conformation $X$ is expressed by the
Hamiltonian $H(X)=\varepsilon(m_{AA}+m_{BB}- m_{AB})$. The quantities
$m_{MM'}$'s count the numbers of contacts between monomers of the kind
$M$ and $M'$, where 
$M,M'=A,B$.
Let
$\bos R_1,\ldots,\bos R_N$ be the position of the monomers.
Two monomers $i$ and $j$ are said to be in contact if
$i\ne j\pm1$ and $|\bos R_i-\bos R_j|=1$.
$\varepsilon$ is the energy cost of one contact, which is positive
when the $A$ and $B$ monomers are in contact with themselves and
negative if a 
monomer of type $A$ is in contact with a 
monomer of type $B$. 
Thermodynamic units will be assumed in which the Boltzmann constant
is equal to one. For convenience, the rescaled Boltzmann
factor $\bos T=\frac T\varepsilon$ and rescaled Hamiltonian $\bos
H(X)=\frac{H(X)}\varepsilon$  are introduced. The new Hamiltonian
$\bos H(X)$ takes only integer values.
The simulations are performed using the Wang-Landau Monte Carlo
algorithm \cite{wl}. The details on the sampling and the treatment of the
topological constraints can be found in
Refs.~\cite{yzff} and \cite{yzff2013}.

The partition function of the polymer knot is 
given by: $Z(\bos T)=\sum_{{\cal E}=m_1}^{m_2}e^{-{\cal E}/\bos
  T}\phi_{\cal E}$, where $\phi_{\cal E}$ denotes the density of
states:
$\phi_{\cal E}=\sum_X\delta(\bos H(X)-{\cal E})$. $\phi_{\cal E}$ is
the quantity to be evaluated via Monte Carlo methods. The integers
$m_1$ and $m_2$ define the energy interval ${\cal I}=[m_1,m_2]$ over which the
sampling is 
performed. 
The
expectation values of any observable $\cal O$ may be computed
using the formula:
$\langle{\cal O}\rangle(\bos T)=\frac 1{Z(\bos T)}\sum_{{\cal
E}=m_1}^{m_2}e^{-{\cal E}/\bos   T}\phi_{\cal E}{\cal O}_{\cal E}$.
\begin{figure}
  \begin{center}
    \includegraphics[width=0.48\textwidth]{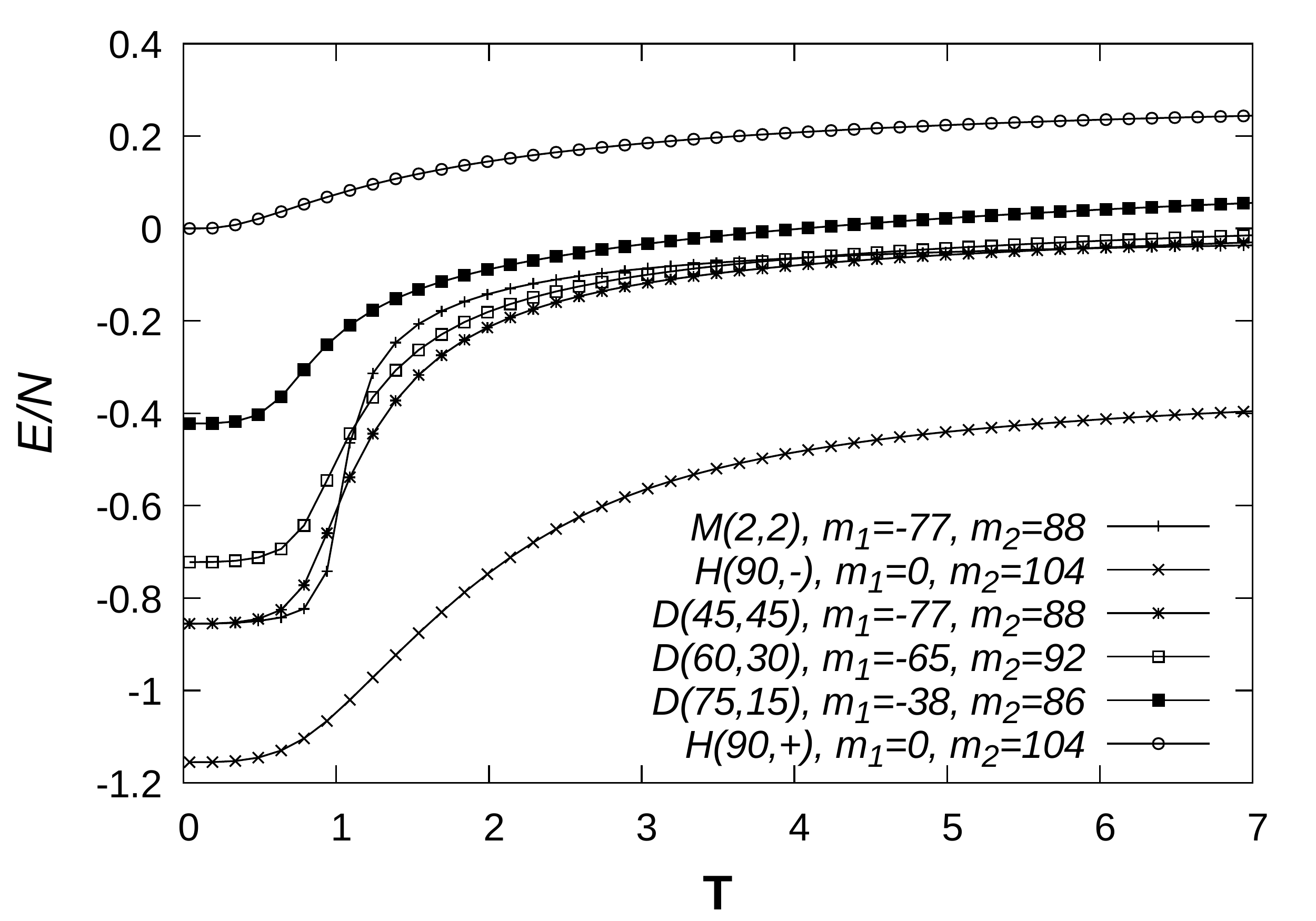}
  \caption{The specific energy of a knot $3_1$ with $N=90$ in various
monomer configurations is plotted as a function of $\bos T$.}\label{fig-local-N90-a}
  \end{center}
\end{figure}
Here ${\cal O}_{\cal E}$ denotes the average of $\cal O$ over all
sampled states with rescaled energy $\cal E$. The observables that
will be considered in this work are the specific energy $\frac {E(\bos
T)}N=\sum_{{\cal E}=m_1}^{m_2}{\cal E}e^{-{\cal E}/\bos
  T}\phi_{\cal E}$, the specific heat capacity $C/N=\frac
1N\frac{\partial E(\bos T)}{\partial \bos T}$ and the mean square
average of the gyration radius $R_G^2$.

\begin{figure}
  \begin{center}
    \includegraphics[width=0.48\textwidth]{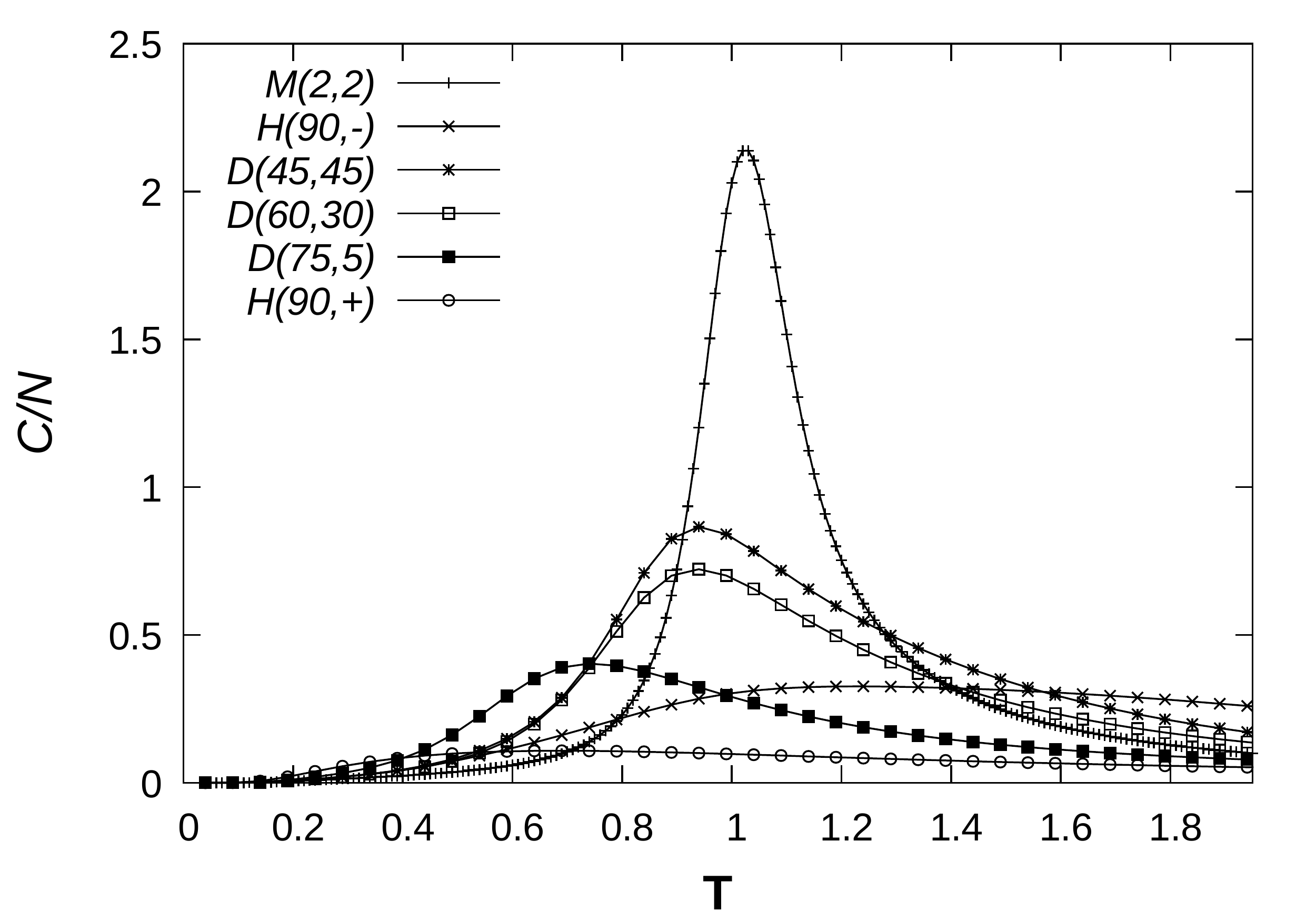}
 \caption{The specific heat capacity of a knot $3_1$ with $N=90$ in various
monomer configurations is plotted as a function of $\bos T$.}\label{fig-local-N90-b}
  \end{center}
\end{figure}

\begin{figure}
  \begin{center}
    \includegraphics[width=0.48\textwidth]{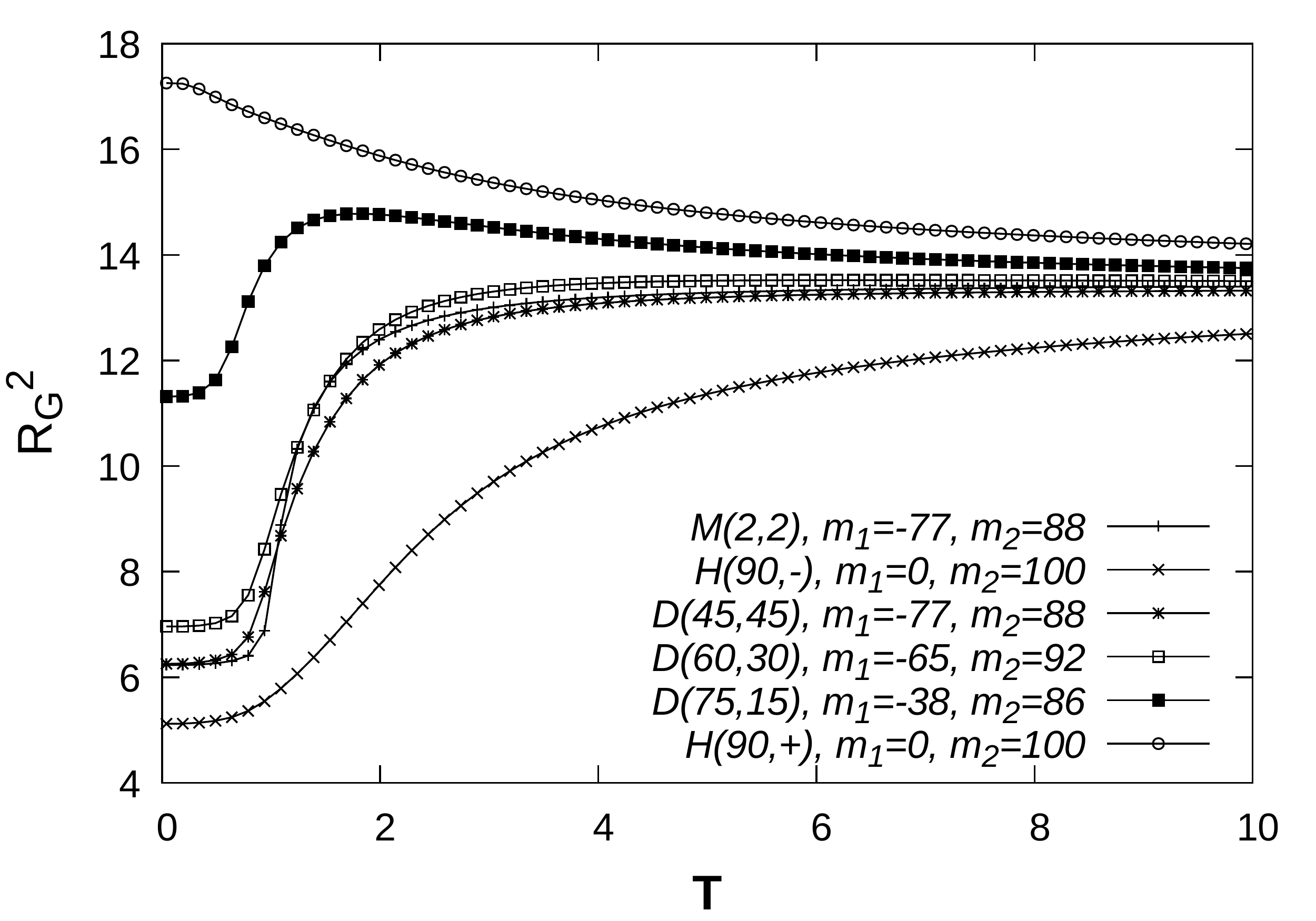}
    \caption{The gyration radii of a knot $3_1$ with $N=90$ in various
monomer configurations is plotted as a function of $\bos T$.}\label{fig-local-N90-c}
\end{center}
\end{figure}

The variety of behaviors that it is possible to obtain in copolymer
knots is shown 
in Figs.~\ref{fig-local-N90-a}--\ref{fig-local-N90-c}. The  same
trefoil knot $3_1$ of length 
$N=90$ is considered and only the  distribution of the $A$ and $B$
monomers is different.
From the plots of the specific energy and of the gyration radius
in 
Figs.~\ref{fig-local-N90-a} and \ref{fig-local-N90-c}
it turns out that the homopolymer trefoil knot $3_1$ has the slowest
swelling rate when the temperature is increasing,
independently if the interactions are repulsive as in the variant
$H(90,+)$ or attractive as in $H(90,-)$.
On the contrary, the swelling process becomes abrupt in 
the multiblock $3_1$ copolymer $M(2,2)$. This knot undergoes a
fast transition from the compact to swollen phase at
$\bos T\sim 1.2$, which causes the marked peak in the specific heat
capacity in Fig.~\ref{fig-local-N90-b}. This feature is present also
in longer knots and persists even
if the size of the blocks is increased, for instance in the case of
$M(8,8)$ copolymers. Another remarkable
phenomenon appears in the plot 
of $R_G^2$ of Fig.~\ref{fig-local-N90-c}. While homopolymers are  simple
systems whose size
steadily increases with growing temperatures,
the $3_1$ copolymer $D(75,15)$ exhibits a more complicated behavior.
Its $R_G^2$ is smallest at low temperatures and 
increases up to its maximum value at intermediate temperatures. After
that, it starts to decrease and finally stabilizes
to some value between the maximum and the minimum at high
temperatures.
These three different
regimes, compact, ultra swollen and swollen are not present in the
 $3_1$ knot with monomer distribution $D(45,45)$.
The distribution of the $A$ and $B$ monomers selects also the allowed
energetic states 
and the range of possible sizes. For example, in
Figs.~\ref{fig-local-N90-a}--\ref{fig-local-N90-c} the $3_1$ copolymer knot
$D(75,15)$  has a lower  energy limit that is bigger than that of its
 counterpart of type $D(45,45)$.

 One goal of this work is to investigate
 how the topology of a knot influences its thermal and mechanical behaviors.
In the case of homopolymers it is known that
topological effects
are relevant especially when the knot is short
and fade out with increasing length, see e.~g. \cite{yzff2013}.
In $D(N_A,N_B)$ copolymers these effects are further enhanced.
The reason is that cyclic $AB-$diblock copolymers
consist of two
homopolymers joined together at both ends. They look locally as
homopolymers, but their global structure emerges 
whenever there are conformations in which
a relevant number of $A$ and $B$ monomers is close to
each other. Since conformations of this kind are
characterized by low energy values, a connection is established between
 the global structure of the knot, which is in turn depending on its
topological properties, and conformations that are
energetically favored.
\begin{figure}
  \begin{center}
    \includegraphics[width=0.48\textwidth]{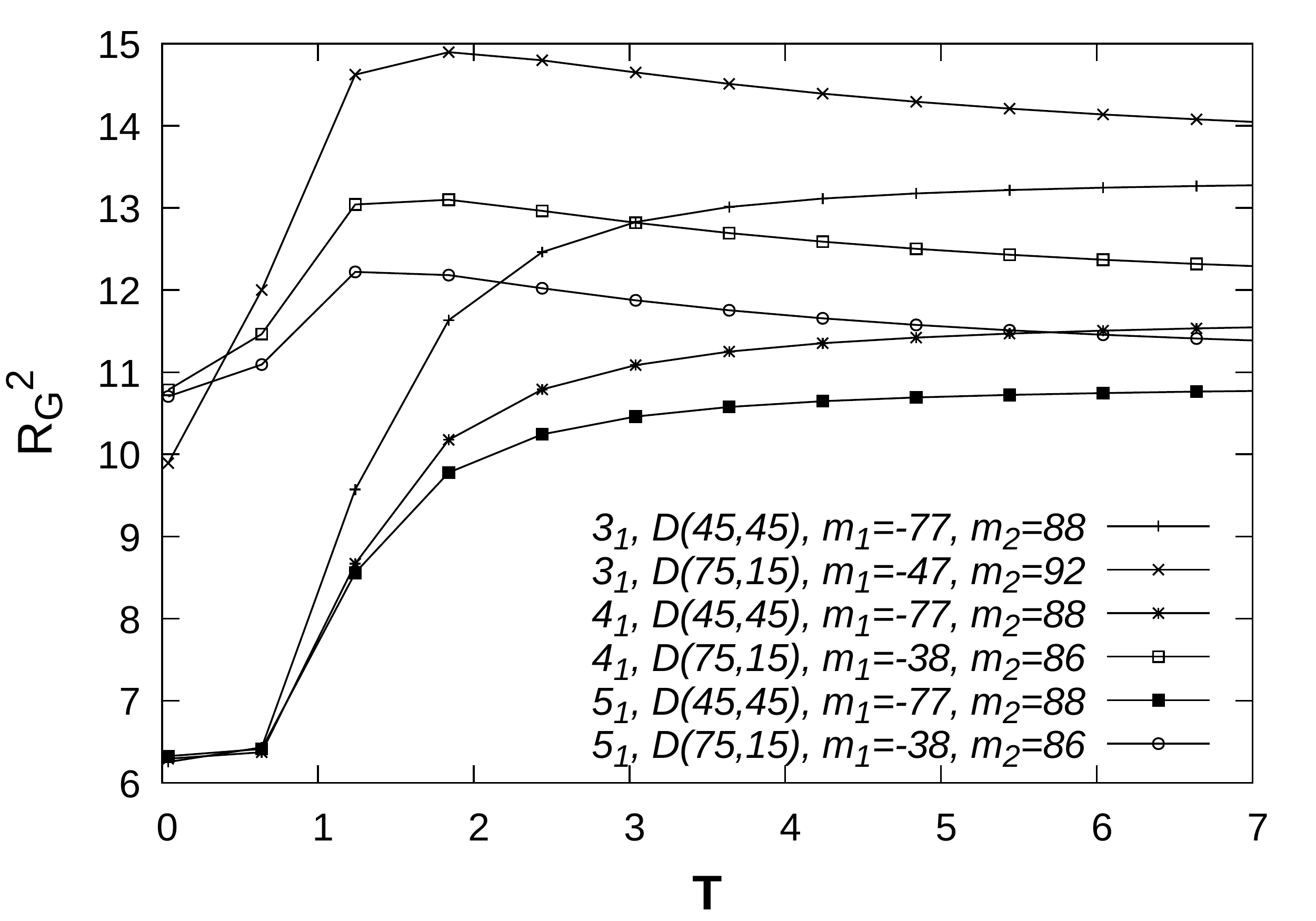}
    \caption{Dependence of the mean square gyration radius of the
knots $3_1,4_1$ and $5_1$ with $N=90$ on the $A/B$ monomer ratio
$\eta$. Two values of $\eta$ are considered: $\eta=1$ and $\eta=5$.}\label{fig-top-N90}
\end{center}
\end{figure}
The plots in Fig.~\ref{fig-top-N90} concerning the gyration radius of
a few knots of copolymer type $D(N_A,N_B)$ show that  strong
topological effects are
at work. For example, the size of the knot $3_1$ changes
substantially with rising temperature in both cases
$\eta=1$ or $\eta=5$.
On the contrary, the size
of the knots
$4_1$ and $5_1$ is bound to vary within a much narrower interval when $\eta=5$.
The compact, ultra swollen and swollen regimes are observed only in the
knots with $\eta=5$.
We remark that in Fig.~\ref{fig-top-N90}
the gyration radii of
knots  differing only by the
value of $\eta$ converge to the same limit at high temperatures.
The same convergence is present also in
Fig.~\ref{fig-local-N90-c}. This is expected 
from the fact that, at high temperatures,
the short-range interactions become irrelevant due to strong thermal
fluctuations.
\begin{figure}
  \begin{center}
    \includegraphics[width=0.48\textwidth]{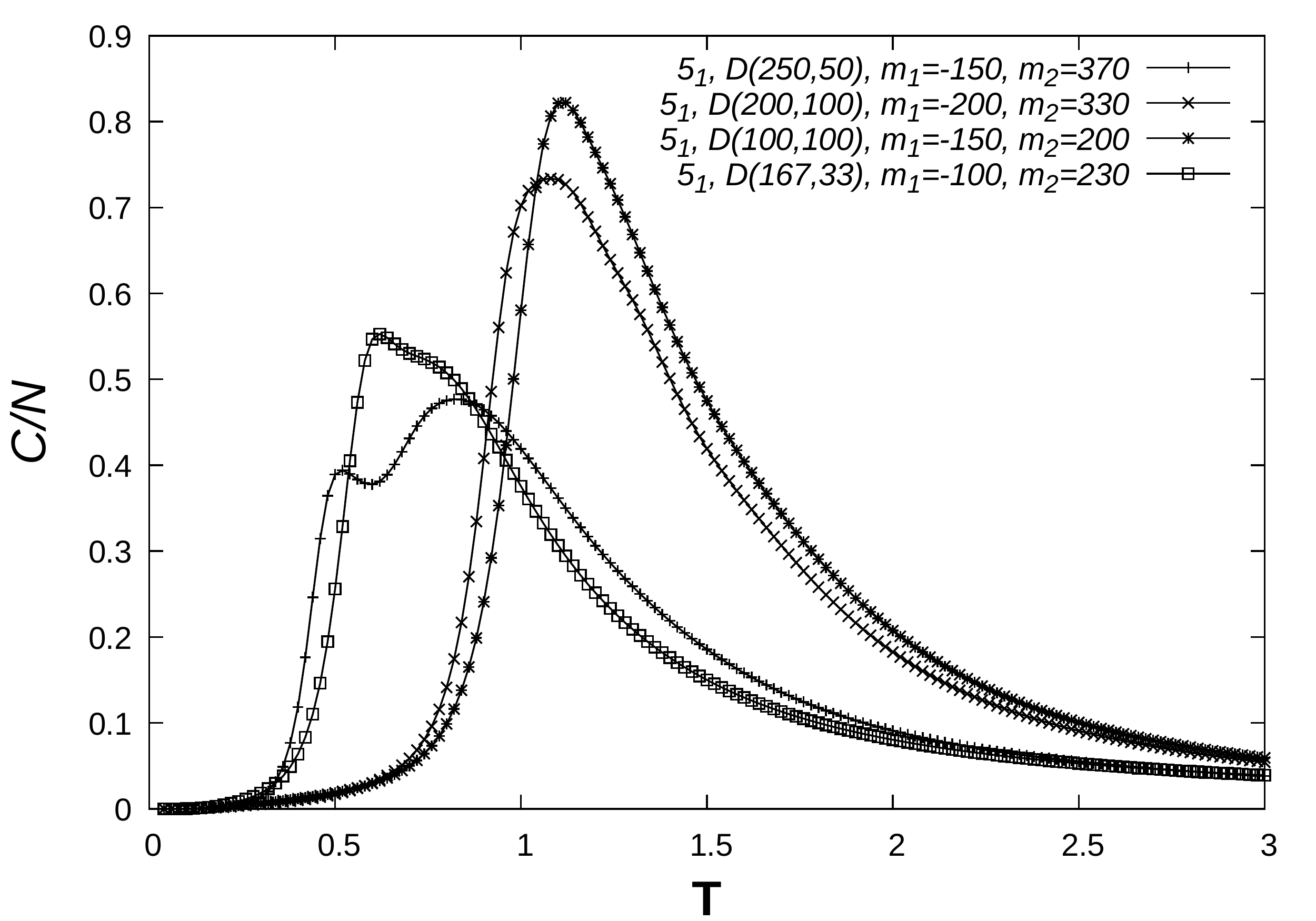}
\caption{Plots of the specific heat capacity of a diblock copolymer
knot $5_1$ with  length $N=300$ in three different monomer
configurations corresponding to  $\eta=1,2,5$.
The fourth plot displays the heat capacity of a knot $5_1$ with
$N=200$ and $\eta=5$.}\label{fig-new-events}  
\end{center}
\end{figure}

As already discussed,  energetically favorable conformations  in which
the $A$ and $B$ monomers are close to each other
are sensitive to the global properties of the knot and thus to
its topology. It is therefore licit to expect the existence of an
interplay between 
 monomer distributions and topology.
To show that, diblock copolymers in the knot configurations $3_1$,
$4_1$, $5_1$ and $9_1$ 
have been studied varying their lengths and  $A/B$ monomer
ratios. The performed analysis points out
that by increasing the knot topology or the
$A/B$ monomer ratio $\eta$,  similar results are produced.
For instance, 
relevant changes in size comparable to those
found in the knot $5_1$ with $N=90$
when passing from $\eta=1$ to $\eta=5$, see Fig.~\ref{fig-top-N90},
can be obtained also
when the length is increased to $N=216$, but only after switching to more
complex knots, such as the knot $9_1$.
In the topologically simpler knot $3_1$ with $N=200$, the size
differences at any given temperature are not so relevant.
The situation changes if in the knot
$3_1$ with $N=200$ the parameter $\eta$ is further increased.
In order to give a rough estimation of the scale of the change, we
take into account the maximum
 value of the mean
square average of the gyration
radius
$R_{G,max}^2$. When
$\eta\sim 7.7$ we find that
$R_{G,max}^2\sim 42.63$. This point of maximum 
is observed at a temperature corresponding to  $\bos T\sim 2.64$.
When  $\eta=1$, instead, the quantity $R_G^2$  steadily increases with rising
temperatures and attains its maximum value of 
$R_{G,max}^2\sim 37.32$ in the upper limit $\bos T= 20.00$ of the
studied interval of temperatures.
The data also indicate  that the fading out of the influence of
topology on the knot behavior is not so fast as in homopolymers. The
details of these calculations will be reported elsewhere.

Another interesting aspect of copolymer knots is related to 
the effects occurring when the length $N$ is increased.  When $N=90$, we
have seen in 
Figs.~\ref{fig-local-N90-a}--\ref{fig-top-N90} that there is 
a swelling phase, followed  by a relevant decrease of the knot size provided
the topological complexity or the $A/B$ monomer ratio are big enough.
Even if the values of $m_1$ and $m_2$ are pushed almost at the limits
of the range of allowed energies, the
specific energy
 of copolymer knots with length $N<200$ and
homopolymer knots up to $N=2100$ is always characterized by a single peak.
The situation is different when longer copolymers are considered at
high values of $\eta$. For example, the knot $5_1$ with $N=300$ and
$\eta=5$ exhibits in Fig.~\ref{fig-new-events} a smaller peak besides
the main one, which corresponds to the swelling process.
This secondary peak appears at very low temperatures and is quite
narrow. It is probably related to the transition from rare, very
compact states. 
In the remaining plots shown in Fig.~\ref{fig-new-events} the
specific heat capacities of a $5_1$ knot
with $N=300$ and $\eta=1,2$  and of a $5_1$ knot with $N=200$ at
$\eta=5$ are shown. No second peak has been 
observed in these two cases.

The
simulations presented in this paper have required the sampling of an
extensive amount of knot conformations. This has been possible thanks
to the introduction of some improvements that have sped up
considerably the original 
Wang-Landau algorithm. 
The major problem to be solved  is that  of rare events, see
\cite{yzff2013} and 
references therein.
Some of the 
conformations
appear after several hundreds of billions of trials and their  inclusion 
would extend enormously the time of the calculations.
The basic idea is to leave open the energy range while requiring the
flatness condition of the energy histogram, which is crucial in the
Wang-Landau algorithm, inside some definite energy interval
${\cal I}=[m_1,m_2]$.
If the sampling process is trapped in some rare conformation outside ${\cal
  I}$, where a precise account of the statistics of the events is not
relevant, it is easy to get rid of this difficulty.
Inside ${\cal
  I}$, instead, the sampling of the conformations, including rare
states, is made faster by introducing a two-step sampling procedure,
in which during the first step the so-called modification factor
becomes a function of the energy.

\begin{figure}[h]
\begin{center}
\includegraphics[width=0.5\textwidth]{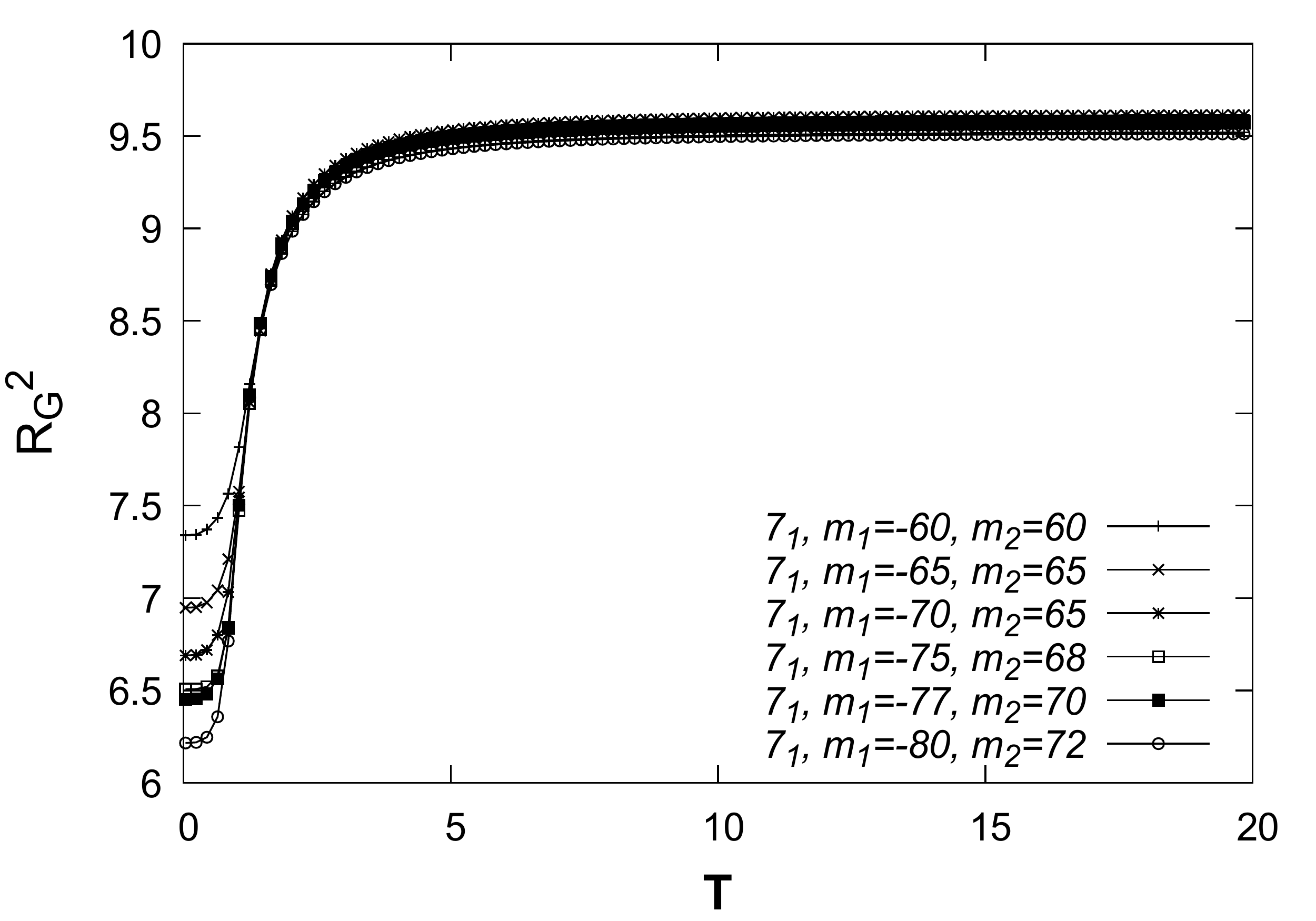}
\caption{The gyration radius $R_G^2$ of the knot $7_1$ with $N=90$  and
  copolymer type $T_1(45,45)$ is plotted for
  increasing widths of the considered energy interval $\cal
  I$. } \label{fig-convergence-tris}   
\end{center} 
\end{figure}

It should be noted that the
expectation values of the
observables are sensitive on the choices of $m_1$ and
$m_2$. The energy cut-off $m_1$ plays a key role
in the study of low temperatures, where low energy conformations are
preferred. At very high temperatures, instead,
the results are quite independent 
of the values of $m_1$ and $m_2$. 
The reason is that at very high temperatures the knot conformations
are swollen and their number of contacts is much higher than $m_1$ and
much lower than $m_2$.
To avoid biases due to a poor choice of the cut-offs $m_1$ and $m_2$ which
eliminates statistically
relevant conformations, the energy interval 
${\cal I}=[m_1,m_2]$ is extended gradually, until
the measured values start to converge to some limit and do not change
substantially after a further extension of ${\cal I}$.
 In Fig.~\ref{fig-convergence-tris} the plots of the
mean square gyration radius
of a knot $7_1$ for increasing energy intervals $\cal I$ are reported.
These plots provide just an example of a trend which is general.
First of all, as previously explained, the results at low
temperatures  are mostly affected by the choice of $m_1$. Secondly,
even if the relatively relevant part of the lower energy spectrum is
neglected, see the plot in the worsest case $m_1=-60,m_2=60$,
 the data at very high temperatures are reproduced with great
precision. Let us note that the measurement of the gyration radius is
particularly 
difficult,
because it is necessary to have a good statistics for each allowed value
of the energy.
The agreement of  the plots analogous to
those of Fig.~\ref{fig-convergence-tris}, but drawn in the case of the
specific energy, is nearly perfect.
It turns out the most important features of the behavior of
a polymer knot are largely independent of the values of $m_1$ and
$m_2$.
For instance, from all diagrams of Fig.~\ref{fig-convergence-tris}
it is possible to obtain a good estimation of the
swelling rate of the knot  with increasing temperatures.
This allows to predict
if the transition
from the compact phase to the swollen phase occurs abrutly in a small
interval of temperatures or slowly over a wider range of temperatures.
Even the lowest resolution plot of the gyration radius in
Fig.~\ref{fig-convergence-tris} allows to conclude that
in
a copolymer
knot $7_1$ of monomer configuration $D(45,45)$ and length $N=90$ the
ultra swollen regime is not present.
\begin{acknowledgments} 
The simulations reported in this work were performed in part using the HPC
cluster HAL9000 of the Computing Centre of the Faculty of Mathematics
and Physics at the University of Szczecin.
The work of F. Ferrari results within the collaboration of the COST
Action TD 1308. The use of some of the facilities of the Laboratory of
Polymer Physics of the University of Szczecin, financed by 
a grant of the European Regional Development Fund in the frame of the
project eLBRUS (contract no. WND-RPZP.01.02.02-32-002/10), is
gratefully acknowledged. 
\end{acknowledgments}


\begin{thebibliography}{99}
\bibitem{sauvage} {\it Molecular Catenanes, Rotaxanes and Knots, A
  Journey Through the World of Molecular Topology}, J. P. Sauvage,
  C. Dietriech-Buchecker (Eds.), (Wiley-VCH Verlag, Weinheim, 1999).
\bibitem{arsuaga} J. Arsuaga, J. Roca and D. W. Sumners,
  {\it Topology of viral DNA}, in {\it Emerging Topics in Physical
Virology },  P. G Stockley and R. Twarock (Eds.), (Imperial College
Press, London, 2010).
\bibitem{ramirez} G. Gil-Ramirez, D. A. Leigh and A. J. Stephens,
{\it Catenanes: Fifty Years of Molecular Links}, {\it Angewandte
Chemie International Edition} {\bf 54} (21) (2015), 6110. 
\bibitem{amabilino} J.-P. Sauvage and D. B. Amabilino. {\it Templated
synthesis of knots and ravels}, in {\it Supramolecular Chemistry: From
Molecules to Nanomaterials}, P. A. Gale and J. W. Steed (Eds.),
(Wiley Online Library, 2012).
\bibitem{tezuka} {\it Topological Polymer Chemistry}, Y. Tezuka (Ed.),
  (World Scientific, Singapore, 2013).
\bibitem{wang} W. Wang, Y. Li  and Z. Lu, {\it Science China Chem.} {\bf
  58} (9) (2015), 1471.
\bibitem{wl}F. Wang and D. P. Landau, Phys. Rev. Lett. 86 (2001), 2050.
\bibitem{yzff} Y. Zhao and F. Ferrari, JSTAT {\it J. Stat. Mech.}
(2012), P11022.
\bibitem{yzff2013}Y. Zhao and F. ferrari, J. Stat. Mech. (2013),
  P10010.
\end{thebibliography}
\end{document}